\documentclass[10pt]{article}

\usepackage{amsmath}
\usepackage{amssymb}

\usepackage{graphicx}
\usepackage{bm}

\usepackage{cite}

\usepackage[cmyk]{xcolor}
\usepackage{soul}
\usepackage{upgreek}

\usepackage{longtable}
\usepackage{cleveref}
\usepackage{ulem}
\crefname{figure}{\textcolor{blue!80!black}{Fig.}}{\textcolor{blue!80!black}{Fig.}}
\crefname{equation}{\textcolor{blue!80!black}{Eq.}}{\textcolor{blue!80!black}{Eq.}}

\usepackage{setspace}
\usepackage{booktabs}
\usepackage[version=3]{mhchem}

\usepackage[labelfont=bf,labelsep=period,justification=raggedright]{caption}

\usepackage[misc]{ifsym}
\usepackage[T1]{fontenc}

\usepackage[super,comma,sort&compress]{natbib}

\newcommand*\pa{\partial}

\makeatletter
\renewcommand{\@biblabel}[1]{\quad#1.}
\makeatother

\date{}

\begin{document}

% Title must be 150 characters or less
\begin{flushleft}
{\Large \textbf{Quantifying energy landscape of high-dimensional oscillatory systems by diffusion decomposition}}
%{\Large \textbf{Quantification of cancer-immunity interplay dynamics by landscape and optimization of transition actions}}
% Insert Author names, affiliations and corresponding author email.
\\
Shirui Bian$^{1,\dagger}$, Ruisong Zhou$^{1,\dagger}$, Wei Lin$^{1,2,4,5,\ast}$, Chunhe Li$^{1,2,3,6,\ast}$
\\
\textbf{1} School of Mathematical Sciences, Fudan University, Shanghai 200433, China
\\
\textbf{2} Shanghai Center for Mathematical Sciences, Fudan University, Shanghai 200438, China
\\
\textbf{3} Institute of Science and Technology for Brain-Inspired Intelligence and MOE Frontiers Center for Brain Science, Fudan University, Shanghai 200433, China
\\
\textbf{4} Shanghai Artificial Intelligence Laboratory, Shanghai 200232, China
\\
\textbf{5} State Key Laboratory of Medical Neurobiology and MOE Frontiers Center for Brain Science, Institute of Brain Science, Fudan University, Shanghai 200032, China
\\
\textbf{6} Lead contact

$\dagger$ These authors contributed equally.\\
$\ast$ Correspondence: wlin@fudan.edu.cn (W.L.), chunheli@fudan.edu.cn (C.L.)
\end{flushleft}

% Please keep the abstract between 250 and 300 words
\section*{Summary}

High-dimensional networks producing oscillatory dynamics are ubiquitous in biological systems. Unravelling the mechanism of oscillatory dynamics in biological networks with stochastic perturbations becomes paramountly significant. Although the classical energy landscape theory provides a tool to study this problem in multistable systems and explain cellular functions, it remains challenging to quantify the landscape for high-dimensional oscillatory systems accurately. Here we propose an approach called the diffusion decomposition of Gaussian approximation (DDGA). We demonstrate the efficacy of the DDGA in quantifying the energy landscape of oscillatory systems and corresponding stochastic dynamics, in comparison with existing approaches. By further applying the DDGA to high-dimensional biological networks, we are able to uncover more intricate biological mechanisms efficiently, which deepens our understanding of cellular functions.

\section*{Introduction}

Many biological systems display periodic oscillatory behaviors, such as cell cycle, circadian rhythm, and population fluctuation \cite{Allada1998Cell,Hastings2018Science,Li2014PNAS,Gerard2009PNAS}. Unveiling the quantitative mechanisms behind the oscillatory behaviors of biological systems holds great scientific significance and practical value \cite{Qin2021nc,Blasius2020Nature,Goldobin2010PRL,Perez2021PRL}. Traditionally, oscillations have been studied using deterministic dynamical models \cite{Gerard2009PNAS,Zxp2011PNAS}. However, stochasticity often plays critical roles in influencing the dynamical behaviors of real-world complex systems \cite{Los2008sci,Kaern2005Nature,Coomer2022CS}. Therefore, exploring the stochastic dynamics and mechanisms of oscillatory systems is crucial and poses great challenges, especially when dealing with high-dimensional biological systems \cite{Li2014PNAS,Qian2012ARB}. Some constructive approaches using phase-amplitude reduction \cite{Goldobin2010PRL,Yoshimura2008PRL} can analyze the stochastic oscillation in the perspective of a reduced coordinate, but it remains challenging to apply this type of approaches to a realistic high-dimensional oscillatory network.

Meanwhile, the energy landscape theory has been proposed to study the global stochastic dynamics of biological systems \cite{Li2014PNAS,Waddington1957,Ge2016PRE,Wang2017Elife,Zhao2024EPR,Lv2015PLoS,Moris2016NPG,Shakiba2022Cell,Zhou2016JCP,Shi2022NSR,Li2020Sci,Frauenfelder1991science}, and provides an important tool to tackle above challenges. To quantify the energy landscape of biological networks, different approaches have been proposed \cite{Zhao2024EPR,Zhou2016JCP,Li2014PNAS,Kang2021Adv}. Among them, an important type of approaches for constructing energy landscape is to solve the Fokker-planck equation (FPE) approximately \cite{Li2014PNAS,Kang2021Adv,Li2016PCCP}. However, high dimensions and nonlinear driving forces often pose great challenges in solving the FPE, even approximately. To address this issue, the weighted summation of Gaussian approximation (WSGA) approach and the extended Gaussian approximation (EGA) approach have been proposed \cite{Kang2021Adv,Bian2023Chaos}, which leverages the Gaussian distributions with proper weight vectors/functions and the moment equation approximation. Both WSGA and EGA approaches are characterized by Gaussian-like diffusion along trajectories, as supported by Kurtz's Central Limit Theorem (CLT) \cite{Kurtz1981AOP}, and use Van Kampen's $\Omega$-expansion \cite{Van2007,Hu1994} to derive the moment equation. Of note, recent works have broadened the theoretical foundations of energy landscape concept, by showing that the landscape of biochemical reaction networks emerges in the mathematical limit $N\to \infty$ \cite{Ge2016PRE,Ge2012}. Variants of approaches rooted in the WSGA have been extensively employed in diverse biological multistable systems \cite{Li2013PLoS,Kang2021Adv,Bian2023Chaos,Shi2022NSR}.

However, oscillatory dynamics bring substantial difficulties to typical approaches of landscape quantification, particularly in handling the moment equations of the WSGA. For instance, introducing oscillation into the WSGA's covariance equations frequently results in an ``explosion'' solution that diverges towards infinity, rendering the original WSGA ineffective. Based on weak correlation assumption, previous works have proposed mean field approximation to avoid this issue \cite{Li2014PNAS}. However, the theoretical explanations for the ``covariance explosion'' remain absent, and it is urgent to develop more accurate and efficient approaches specifically for oscillatory systems.

In this study, we present an effective approximation approach, named as diffusion decomposition of Gaussian approximation (DDGA), for quantifying the landscape of periodic oscillatory systems. Through decomposing the noise to tangential and normal directions, we solve the problem of covariance explosion. To illustrate the advantages of the DDGA, we compare it against two established methods, the WSGA and the EGA with weak correlation assumption, using various networks with different dimensions. Our findings reveal that the DDGA offers improved precision and higher efficiency in quantifying the landscape of high-dimensional oscillatory systems. When applied to complex cell cycle networks, the DDGA reveals more intricate mechanism of biological functions, which is supported by single cell experimental data and is indiscernible using existing approaches.

\section*{Results}

\subsection*{Preliminary formulation}

We first review the WSGA method for approximating the probability distribution. In a $n$-dimensional stochastic dynamical system $\frac{\text{d}\boldsymbol{x}}{\text{d}t}=\boldsymbol{F}(\boldsymbol{x})+\boldsymbol{G}(\boldsymbol{x})\boldsymbol{\Gamma}(t)$, the FPE (in Ito sense) is often used to characterize the distribution $\rho(\boldsymbol{x},t)$ as \cite{Hu1994,Risken1996Fokker}
\begin{equation}\label{FPEn}
	\begin{aligned}
		\frac{\pa\rho (\boldsymbol{x},t)}{\pa t}=&-\sum\limits_{i=1}^{n} \frac{\pa}{\pa x_i}\Big[F_i(\boldsymbol{x})\rho(\boldsymbol{x},t)\Big]+D \sum\limits_{i=1}^{n} \sum\limits_{j=1}^{n} \frac{\pa^2}{\pa x_i\pa x_j}\Big[d_{ij}(\boldsymbol{x})\rho(\boldsymbol{x},t)\Big],\\
		d_{ij}(\boldsymbol{x})=&\sum\limits_{k=1}^n G_{ik}(\boldsymbol{x})G_{kj}(\boldsymbol{x}),
	\end{aligned}
\end{equation}
where $\boldsymbol{F}=(F_1,\dots,F_n)^\top$ represents the drift force, $\boldsymbol{G}=(G_{ij})_{1\leq i,j\leq n}$ represents the diffusion force, $D$ is the diffusion coefficient, and $\boldsymbol{\Gamma}=(\Gamma_1,\dots,\Gamma_n)^\top$ is the $n$-dimensional Gaussian white noise satisfying $\langle  \Gamma_i(t),\Gamma_j(s)\rangle=2D\delta_0(t-s)\delta_{ij}$. Here, $\delta_0$ is the Dirac function, $\delta_{ij}$ equals $1$ if $i=j$ and equals 0 otherwise. The FPE provides a precise description for stochastic dynamics but is hard to solve as $\boldsymbol{F}$ is nonlinear. Thus, it is necessary to provide an approximation of the solution. In the previous work, we have demonstrated that the WSGA provides an efficient approach for approximating the landscape of multistable systems \cite{Bian2023Chaos}. In WSGA, we use different Gaussian distributions to provide a local approximation, where each of the Gaussian distribution is uniquely characterized by the first two moment equations \cite{Hu1994,Bian2023Chaos}:
\begin{equation}\label{eq-WSGA}
	\begin{cases}
		\begin{aligned}
            \dot{\boldsymbol{\mu}}(t)=&\boldsymbol{F}(\boldsymbol{\mu}(t)),
            \end{aligned}\\
		\begin{aligned}
			\dot{\boldsymbol{\Sigma}} (t) =& \boldsymbol{\Sigma}(t) \boldsymbol{A}^\top(t) + \boldsymbol{A}(t) \boldsymbol{\Sigma}(t)+D\Big[ \boldsymbol{d} (\boldsymbol{\mu}(t)) + \boldsymbol{d}^\top (\boldsymbol{\mu}(t))\Big].
		\end{aligned}
	\end{cases}
\end{equation}
Here, the elements of the Jacobi matrix $\boldsymbol{A}(t)$ are specified as $\boldsymbol{A}_{(i,j)}(t)=\frac{\partial F_i(\boldsymbol{\mu}(t))}{\partial x_j(t)}$, $\boldsymbol{d}(\boldsymbol{x})=(d_{ij}(\boldsymbol{x}))_{1\leq i,j\leq n}$, $\boldsymbol{\mu}$ and $\boldsymbol{\Sigma}$ are the expectation and the covariance of a Gaussian distribution. The WSGA provides a proper weighted function for these distributions to construct a global distribution and the corresponding energy landscape. Specifically, for multistable systems, we describe the distribution with $N$ stable states by $\textstyle{\rho(\boldsymbol{x},t)=\sum_{i=1}^N \omega_i\times \mathcal{N}(\boldsymbol{\mu}_i(t),\boldsymbol{\Sigma}_i(t))(\boldsymbol{x})}$. Here, $\omega_i$ is the corresponding weight for the $i$-th stable state, and both $\boldsymbol{\mu}_i$ and $\boldsymbol{\Sigma}_i$ provide the moment information (expectation and covariance) of the Gaussian distribution $\mathcal{N}(\boldsymbol{\mu}_i(t),\boldsymbol{\Sigma}_i(t))$ \cite{Bian2023Chaos} (see also Methods S1, S2, and S7).

From \cref{eq-WSGA}, the second-moment equations are critical for approximating the probability distribution. Although the WSGA-based approaches show great performance in multistable systems, they usually fail to cope with oscillatory systems. In many periodic systems, we find that the covariance $\boldsymbol{\Sigma}$ ``explodes'' to infinity as $t\rightarrow +\infty$ (see Note S1), leading to low efficacy of the WSGA in oscillatory systems. Therefore, generally effective mathematical tools are still lacking for studying the stochastic dynamics of oscillatory systems so far.

\subsection*{Approximating landscape using pre-solution and diffusion decomposition}

Here, we propose an approximation approach that addresses the limitations of the original WSGA while retaining its simplicity and effectiveness. Through several examples, we demonstrate the effectiveness and efficiency of this approach for investigating the stochastic dynamics of oscillatory systems.

The original WSGA approach incorporates the idea of ``local diffusion of individuals along noiseless trajectories'' from CLT. While this approach has limitations in global exploration, it still provides us with valuable insights. For simplicity, we hope to illustrate the idea by assuming that the diffusion/noise level are homogeneous and constant, i.e., $\boldsymbol{d}(\boldsymbol{x})= I_n$, although our approach applies for more general systems with a general noise form (see details in Note S3). For the noise $\boldsymbol{\Gamma}\sim \mathcal{N}(0,I_n)$, we decompose it orthogonally along the tangential direction $v_1$ of the limit cycle, and along the ($n-1$)-dimensional normal plane as $\boldsymbol{\Gamma} = \boldsymbol{\Gamma}_t+\boldsymbol{\Gamma}_n$, where $\boldsymbol{\Gamma}_t \sim \mathcal{N}(0,v_1v_1^\top),\, \boldsymbol{\Gamma}_n\sim \mathcal{N}(0,I_n-v_1v_1^\top)$, and $\boldsymbol{\Gamma}_t \perp \boldsymbol{\Gamma}_n$.

We firstly only consider the noise component along the tangential direction and obtain a restricted one-dimensional dynamics on the limit cycle specified by $\nabla\cdot (\boldsymbol F(\boldsymbol x)q(\boldsymbol x)-D\nabla q(\boldsymbol x))=0$ for $\boldsymbol x\in LC$, where $LC$ represents the limit cycle. It is well-known that the one-dimensional FPE has an explicit solution, which we refer to as the ``pre-solution'' of the DDGA (see details in Note S2). It represents the one-dimensional distribution restricted to the limit cycle, up to a multiplicative constant, characterizing the majority of the information related to the oscillatory dynamics. The construction of the pre-solution highly improves the approximation accuracy.

The other key point is the diffusion process on the remaining ($n$-1)-dimensional normal plane. Here, the extent of diffusion is primarily determined by the covariance equations. Under the assumption of small noise level, we demonstrate that the covariance on the normal plane of any point on the limit cycle can be approximated as a constant. In this case, we only need to solve $(n-1)^2$ linear equations instead of solving $(n-1)^2$ ODEs, which significantly reduces the computational complexity. Specifically, the original $n$-dimensional covariance matrix is obtained as $\boldsymbol{\Sigma} = Q\bar{\boldsymbol{\Sigma}} Q^\top + Dv_1v_1^\top$, where $v_1(\boldsymbol{x}^*)$ is the unit tangential vector for $\boldsymbol{x}^*\in LC$, $Q(\boldsymbol{x}^*)=(v_2(\boldsymbol{x}^*),\dots,v_n(\boldsymbol{x}^*))$ represents the transformation matrix from the $(n-1)$-dimensional normal plane basis to the $n$-dimensional basis, and $\bar{\boldsymbol{\Sigma}}$ is the $(n-1)$-dimensional covariance matrix solved through the Lyapunov equations $2D I_{n-1} +\bar{\boldsymbol{\Sigma}}(\boldsymbol{x}) Q(\boldsymbol{x})^\top{\boldsymbol{A}(\boldsymbol{x})}^\top Q(\boldsymbol{x})+ Q(\boldsymbol{x})^\top{\boldsymbol{A}(\boldsymbol{x})}Q(\boldsymbol{x}) \bar{\boldsymbol{\Sigma}}(\boldsymbol{x})=0$ (see detailed derivations and pseudocode in Notes S2 and S4).

Based on the above idea, we improved the WSGA approach and proposed DDGA. Instead of using constant functions describing the weight as in WSGA, the weight in DDGA is derived from the ``pre-solution'' that incorporates information from the periodic dynamics. The updated density function takes the form:
\begin{equation}\label{}
	\begin{aligned}
		\rho(\boldsymbol{x})={\int_{LC}\frac{q(\boldsymbol{z})}{\int_{LC} q(\boldsymbol{y})\text{d}\boldsymbol{y}}\cdot\frac{\text{e}^{-\frac{1}{2}(\boldsymbol{x}-\boldsymbol{z})^\top \boldsymbol{\Sigma}(\boldsymbol{z})^{-1}(\boldsymbol{x}-\boldsymbol{z})}}{(2\pi)^{\frac{n}{2}}\sqrt{\det{\boldsymbol{\Sigma}(\boldsymbol{z})}}}\text{d}\boldsymbol{z}}.
	\end{aligned}
\end{equation}
Correspondingly, the energy landscape $U$ is constructed using $U=-\log \rho_{ss}$, where $\rho_{ss}$ is the steady state distribution \cite{Li2014PNAS,Kang2021Adv}.

To investigate the performance of the DDGA in realistic biological systems, we compare it with the other representative methods, applying to different stochastic systems with various dimensions, including a 6-dimensional synthetic oscillatory network and a 44-dimensional human mammalian cell cycle. Similar results for the Stuart-Landau oscillator (under two sets of parameters) with Gaussian white noise and a 2-dimensional chemical Langevin equation with intrinsic noise are displayed in Notes S6 and S7 (\textcolor{blue!80!black}{Figs.}\,S1, S2, S3 and \textcolor{blue!80!black}{Tab.} S1). These results indicate that the DDGA provides a more accurate and efficient approach for quantifying energy landscape of high-dimensional oscillatory systems under various noise conditions, and will not be affected by different parameter sets.

\subsection*{Application to a synthetic oscillatory network}

\begin{figure}[t!]
	\centering
	\includegraphics[width=0.9\textwidth]{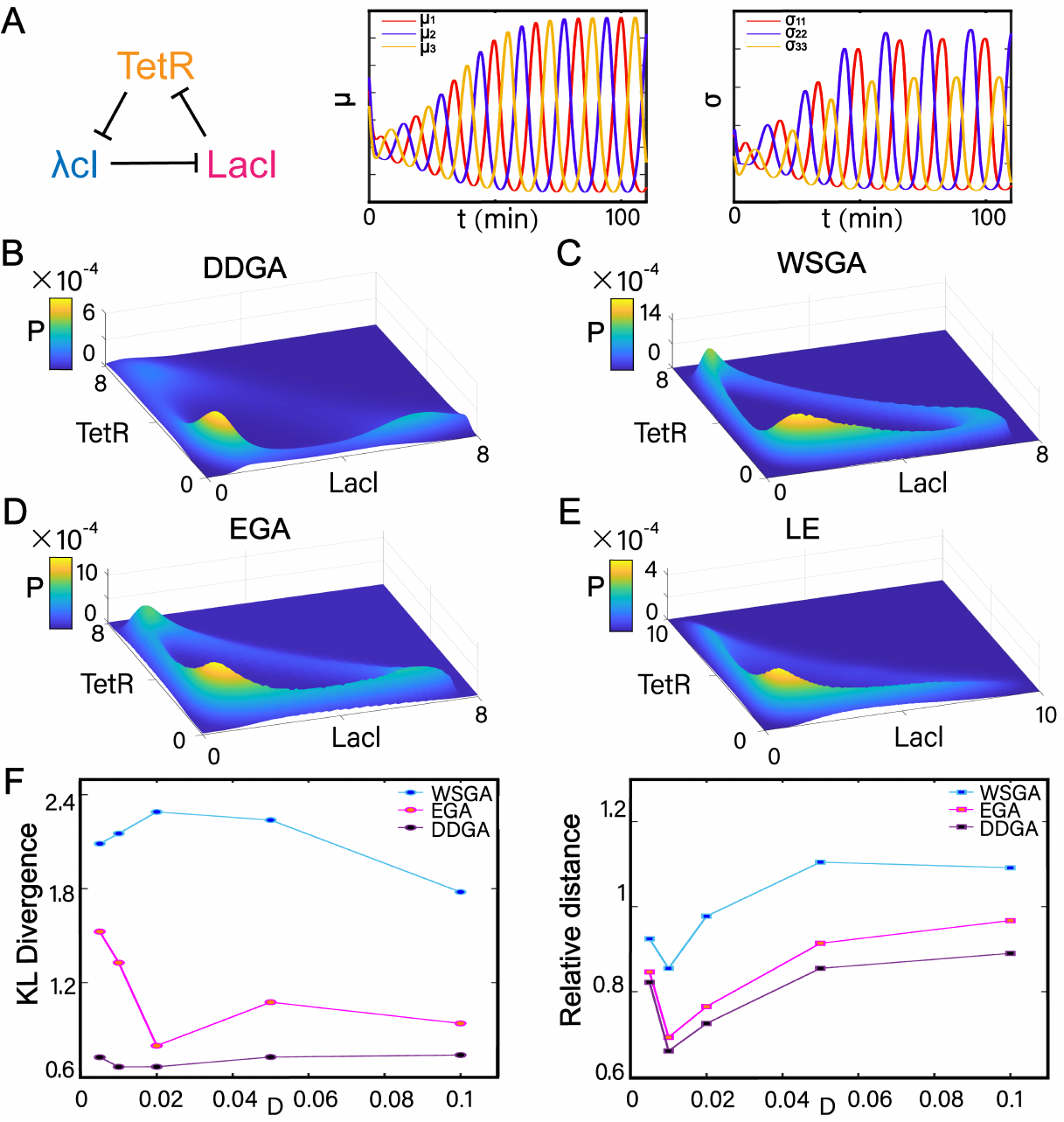}
	\caption{Landscape of synthetic oscillatory network and comparisons using different methods. (A) The negative feedback loop consisting of three repressilators and their oscillatory moments: the expectation $\mu_i$ and the covariance $\sigma_{ij}$. (B-E) Landscape obtained from (B) the DDGA, (C) the WSGA, (D) the EGA, and (E) the LE. (F) Performance of these three approximation approaches in terms of K-L divergence and relative distance. See also \textcolor{blue!80!black}{Fig.} S4 and \textcolor{blue!80!black}{Tab.} S2.}
	\label{Fig_SON1}
\end{figure}

A typical biological oscillator is the synthetic oscillatory network, which has been engineered in Escherichia coli using three transcriptional repressor systems (\cref{Fig_SON1}A) \cite{EM2000}. For the repressilators TetR, LacI, $\lambda$cI, $m_i\,(i=\text{LacI},\,\text{TetR},\,\lambda \text{cI})$, their mRNA concentrations, and $p_i$, their corresponding protein concentrations, are governed by the following stochastic differential equations with corresponding white noise $\Gamma$:
\begin{equation*}\label{}
	\begin{cases}
		\begin{aligned}
			\frac{\text{d} m_i}{\text{d} t}&=-m_i+\frac{\alpha}{1+p_j^n}+\alpha_0+\Gamma_{m_i},(i,j)=(\text{LacI},\lambda \text{cI}),\,(\text{TetR},\text{LacI}),\,(\lambda \text{cI},\text{TetR}),
		\end{aligned}\\
		\begin{aligned}
			\frac{\text{d} p_i}{\text{d} t}=-\beta (p_i-m_i)+\Gamma_{p_i}\,,\,i=\text{LacI},\,\text{TetR},\,\lambda \text{cI},
		\end{aligned}
	\end{cases}
\end{equation*}
where the parameter explanations are listed in \textcolor{blue!80!black}{Tab.} S2. Here, we apply the methods of DDGA, EGA, and WSGA to this system (\textcolor{blue!80!black}{Figs.}\,\ref{Fig_SON1} B-D, and \textcolor{blue!80!black}{Fig.} S4), individually. Meanwhile, the Langevin equation (LE) simulation results (\cref{Fig_SON1}E) are treated as the ground truth solution of the FPE. Following the DDGA procedure, the moments oscillate periodically (\cref{Fig_SON1}A), while the probability looks like a Mexican hat with a triangle-like limit cycle. The DDGA significantly outperforms the WSGA and the EGA in terms of the accuracy and the computational cost (see \textcolor{blue!80!black}{Tab.} S3).

\begin{figure}[t!]
	\centering
	\includegraphics[width=0.9\textwidth]{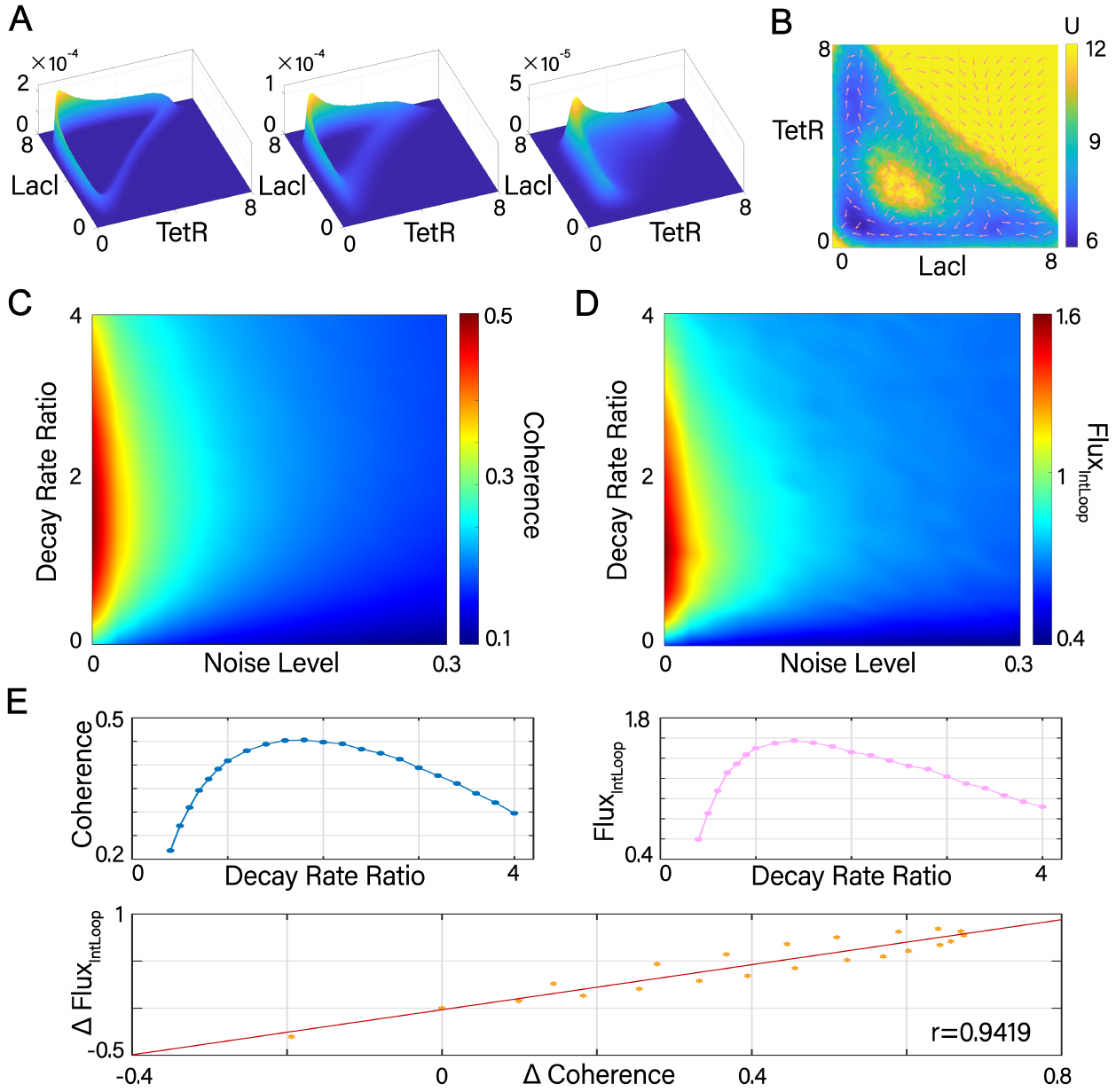}
	\caption{Coherence and flux quantify the oscillation stability of repressilator. (A) The explosion phenomenon of probability reflected by the Langevin simulation, with the noise level increasing from $D=0.002$ (left panel) to $D=0.02$ (middle panel) and $D=0.2$ (right panel). (B) The landscape and the flux for the 2-dimensional projection spanned by the concentration of these repressilators when $D=0.02$. The arrows represent the flux, while the color of the landscape represents the potential at these points. (C-D) The change of the coherence and the flux integration when the noise level and the ratio of decay rate change. (E) The change of two indices when the noise level is fixed as $D=0.02$. Their trends are consistent described by Pearson correlation coefficient $r=0.9419$. See also \textcolor{blue!80!black}{Figs.} S5 and S6.}
	\label{Fig_SON2}
\end{figure}

We also calculate the Kullback-Leibler (K-L) divergence $d_\textbf{KL}(p_r(\boldsymbol{x})\vert p_a(\boldsymbol{x}))=\int p_r(\boldsymbol{x}) \log \frac{p_r(\boldsymbol{x})}{p_a(\boldsymbol{x})}\text{d}\boldsymbol{x}$ and the relative distance $d_\textbf{RD}(p_r(\boldsymbol{x}),p_a(\boldsymbol{x}))=\frac{\int (p_r(\boldsymbol{x})-p_a(\boldsymbol{x}))^2\text{d}\boldsymbol{x}}{\int p_r(\boldsymbol{x})^2\text{d}\boldsymbol{x}}$  to measure the distance for the distribution calculated by the methods of the DDGA, the WSGA and the EGA deviating from the real solution (\cref{Fig_SON1}F, see also Methods S3, S4). Here, we take the probability distribution $p_r(\boldsymbol{x})$, obtained by the LE as the real solution, and $p_a(\boldsymbol{x})$ as the approximating probability distribution calculated from these three approximation methods, respectively. By pre-solution, the DDGA performs much better than the WSGA and the EGA under small noise. This is because under small noise, the diffusion effect is small. Thus, the landscape almost surrounds the limit cycle, which makes it almost directly related to the sub-distribution on the limit cycle, i.e., the pre-solution. Simultaneously, from simulations we discover that, when the diffusion coefficient $D$ becomes larger, the ``Mexican-hat-like'' landscape loses its shape, and the probability ``explodes'' into the inside area of the limit cycle (\cref{Fig_SON2}A). Practically, this explosion phenomenon causes the inefficacy of almost every approximation approach, because none of them can predict the disappearance of the limit cycle. However, our pre-solution leads the DDGA to partially capture this abnormal phenomenon (\cref{Fig_SON1}B).

To further explore the explosion phenomenon, which is possibly caused by the system losing stability, we analyze the stability of this system. Firstly, we display the landscape and the flux for the 2-dimensional projection (\cref{Fig_SON2}B). The flux guarantees the cycling along the limit cycle and maintains the oscillation of the network. Then, we seek to study the effects of key parameters on the robustness of the system. Here we chose two parameters, the noise level $D$ and the decay rate ratio $\beta$, and introduce two indicators, coherence and flux integration \cite{Sasai2007JCP,Li2014PNAS}, to evaluate the stability of the system across parameter variations (see Method S6). We computed the two indicators for this synthetic oscillatory network across various noise levels and decay rate ratios (\textcolor{blue!80!black}{Figs.}\,\ref{Fig_SON2}C, D). Both indicators consistently demonstrate a clear trend with changing parameters. Firstly, as the noise level increases, both indicators decrease, suggesting that the system becomes unstable. Secondly, both indicators exhibit a pattern of increasing to a peak at $\beta\approx 1.8\pm 0.2$ and then decreasing with higher $\beta$. This demonstrates that stable oscillation requires a middle level of decay rate ratio. Similar results under distinct noise levels are shown in \textcolor{blue!80!black}{Fig.} S5.

We also explore the phase transition and the stability of the landscape under different parameter choices using DDGA (\textcolor{blue!80!black}{Fig.} S6). The landscape under a small decay rate ratio exhibits a longer period, with the diffusion phenomena most pronounced at an intermediate level of the decay rate ratio. Furthermore, we find that the two indicators show a highly positive correlation (\cref{Fig_SON2}E) when $D$ is fixed. This arises from the fact that the flux also contributes to the stability of oscillations: Flux plays a crucial role in pushing the oscillation to proceed cyclically along the limit cycle orbit. A stronger flux leads to a more robust cycling inertia, contributing to higher stability of oscillations \cite{Ye2021JCP}.

\subsection*{Application to a high-dimensional mammalian cell cycle network}
\begin{figure*}[t!]
	\centering
	\includegraphics[width=1.0\textwidth]{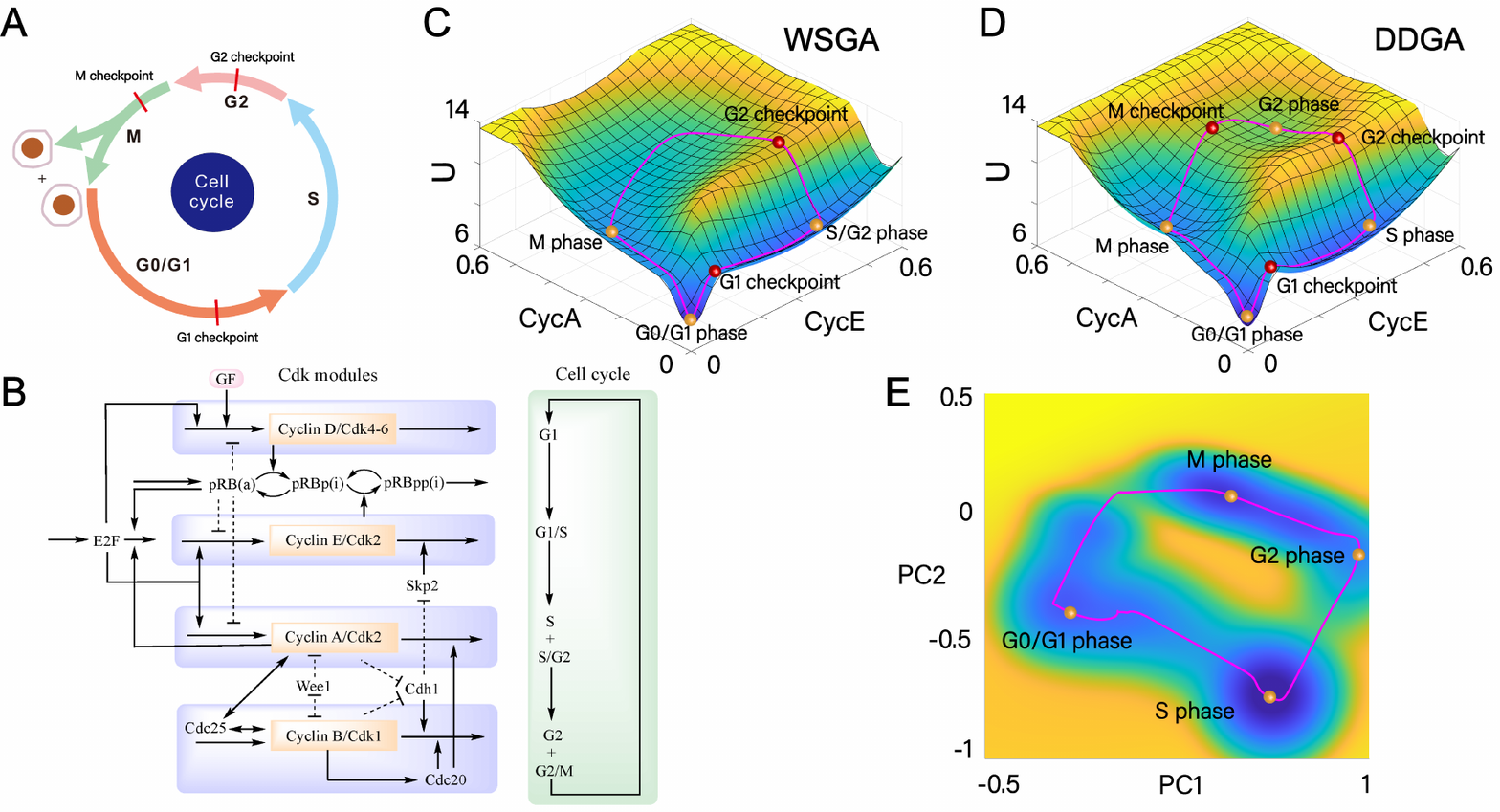}
	\caption{Landscape of the mammalian cell cycle network. (A) An illustration for complete progression of cell cycle in mammalian cells. (B) The wiring diagram of the cell cycle network. The network includes four major cyclin/Cdk complexes controlling the four stages of cell cycle. (C-D) The landscape of the cell cycle network constructed by (C) the WSGA and (D) the DDGA. (E) The landscape is shown in the reduced dimensions using DDGA. See also \textcolor{blue!80!black}{Figs.} S8, S9 and \textcolor{blue!80!black}{Tabs.} S4, S5.}
	\label{Fig_MCC}
\end{figure*}

We further apply the DDGA and the WSGA to a 44-dimensional mammalian cell cycle model \cite{Gerard2009PNAS,Li2014PNAS}, since the EGA is almost ineffective due to the dimensional curse here. The cell cycle model is mainly determined by four central cyclin/Cdk complexes (\cref{Fig_MCC}B) which induce four cell cycle stages (G0/G1, S, G2, and M, \cref{Fig_MCC}A) \cite{Gerard2009PNAS,Gerard2012PLoS}. The dynamics of this model is governed by 44-dimensional ODEs based on the Michaelian kinetics (see Note S13).

Previously we have proposed that on the landscape obtained by the WSGA, the emergence of three local basins and two saddle points can reflect the different phases of the cell cycle progression and the checkpoints of phase transition \cite{Li2014PNAS}. Here, we choose GF=0.5 where the system oscillates, and the diffusion coefficient $D=0.1$ to test whether the DDGA improves the WSGA (see other parameter values and explanations in \textcolor{blue!80!black}{Tab.} S4). The results are shown in \textcolor{blue!80!black}{Figs.}\,\ref{Fig_MCC}C, D. The landscape resulting from the WSGA exhibits three basins that correspond to the G0/G1, the S/G2, and the M phases, along with two checkpoints: the G1 checkpoint and the G2 checkpoint. 

Some essential differences emerge on the landscape using the DDGA. An additional basin associated with the G2 phase is identified, and a new saddle point representing the M checkpoint, emerges along the cycle. It is known that the mammalian cell cycle possesses four stages and three checkpoints. Therefore, the results from the DDGA are more consistent with experiments, and provide more accurate and valid landscape quantification. This fact also holds when we combine the DDGA with the dimension reduction technique to get a landscape in reduced dimensions \cite{Kang2021Adv} (\cref{Fig_MCC}E, see also Method S5 and \textcolor{blue!80!black}{Tab.} S5).

As intriguingly suggested by the computational results, the experimental data from single-cell RNA sequencing exhibit excellent alignment with the landscape generated using DDGA \cite{Capolupo2022Science, Riba2022NC, Qiu2022Cell} (details provided in Note S12). Specifically, the potential wells corresponding to the G0/G1, S, and M phases on the landscape closely match the single-cell data (\textcolor{blue!80!black}{Figs.} S8, S9). This concordance is particularly striking, as it connects theoretical model predictions with experimental observations. Thus, our model-based approach, which captures key features of biological systems, proves robust against the complexities and noise inherent in the data. Such consistency demonstrates the potential of DDGA as a powerful tool to enhance our understanding of real-world complex systems, such as the cell cycle.
\begin{figure}[t!]
	\centering
	\includegraphics[width=1.0\textwidth]{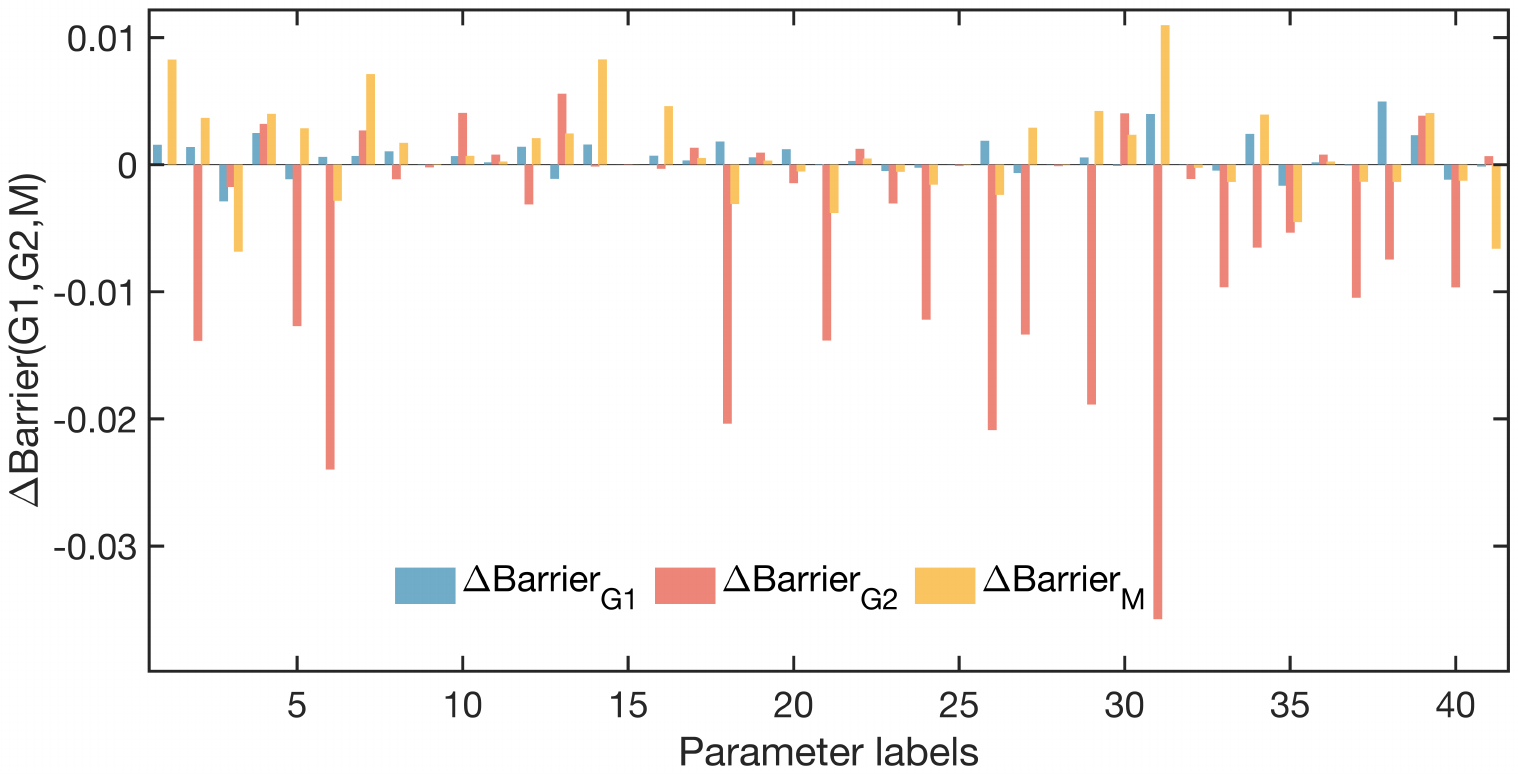}
	\caption{Global sensitivity analysis in terms of the barriers ($Barrier_{\text{G1}}$ for G1 checkpoint, $Barrier_{\text{G2}}$ for G2 checkpoint, and $Barrier_{\text{M}}$ for M checkpoint, see Note S11). The horizontal labels (1 to 41) are corresponding to the 41 parameters (synthesis rate or regulation strength for: ``1: AP1'', ``2: E2F'', ``3: pRB'', ``4: synthesis of Cdc25 acting on CycE/Cdk2'', ``5: Skp2'', ``6: Cdh1'', ``7: synthesis of Cdc25 acting on CycA/Cdk2'', ``8: p27 synthesis independent of E2F'', ``9: p27 induced by E2F'', ``10: CycB'', ``11: Cdc20'', ``12: synthesis of Cdc25 acting on CycB/cdk1'', ``13: Wee1'', ``14: CycD induced by AP1'', ``15: CycD induced by E2F'', ``16: CycD binding Cdk4,6'', ``17: CycD/Cdk4,6 binding P27'', ``18: CycE induced by E2F'', ``19: CycE/Cdk2 binding p27'', ``20: CycE binding Cdk2'', ``21: CycA binding Cdk2'', ``22: CycA/Cdk2 binding p27'', ``23: CycB/Cdk1 binding p27'', ``24: CycB binding Cdk1'', ``25: inhibition of pRB to CycD'', ``26: CycE inhibited by pRB'', ``27: CycA inhibited by pRB'', ``28: p27 inhibited by pRB'', ``29: Cdc25 activating CycA/Cdk2'', ``30: Wee1 inhibiting CycA/Cdk2'', ``31: CycA/Cdk2 activating Cdc25'', ``32: CycA/Cdk2 and CycB/Cdk1 inhibiting Cdh1'', ``33: CycE/Cdk2 inhibiting p27'', ``34: Cdc25 activating CycB/Cdk1'', ``35: Wee1 inhibiting CycB/Cdk1'', ``36: Cdc20 (inactivate form)'', ``37: CycB/Cdk1 activating Cdc20'', ``38: CycB/Cdk1 activating Cdc25'', ``39: CycB/Cdk1	inhibiting Wee1'', ``40: CycE/Cdk2 activating Cdc45'', ``41: ATR activating Chk1''). We calculate the relative barrier change $\Delta Barrier(\text{G1},\text{G2},\text{M})$ of each barrier under 0.1\% perturbation of parameters (each parameter is increased 0.1\% individually). See also \textcolor{blue!80!black}{Fig.} S7 and \textcolor{blue!80!black}{Tab.} S6.}
	\label{Fig4}
\end{figure}

Moreover, to unravel the key determinants influencing the cell cycle, we conducted a global sensitivity analysis on the oscillation period and the barrier height from landscape topography at various checkpoints (\cref{Fig4}, \textcolor{blue!80!black}{Fig.} S7 and \textcolor{blue!80!black}{Tab.} S6, see details in Note S11). Our findings exhibit consistency with previous results while offering additional insights, notably highlighting the M checkpoint, which was not captured in earlier studies \cite{Li2014PNAS}. Most of these new insights agree with experimental evidences.  For instance, our results show that activating Cdc25 (parameters 7 and 31) and pRB (parameter 3) both have a significant effect on $Barrier_{\text{M}}$. These results highlight the crucial role of Cdc25 and pRB in regulating M phase switching, consistent with their importance in mitosis as demonstrated in previous studies\cite{Lici15, 86cell, Longworth2008GD, Coschi2010GD}. Additionally, the activation of Cdh1 is associated with an increase in $Barrier_{\text{G1}}$ and a decrease in $Barrier_{\text{G2}}$ (parameter 6), indicating the roles of Cdh1 on suppressing the G1-S phase transition and promoting the G2 DNA-damage checkpoint crossing. This aligns with previous findings that Cdh1 stabilizes the G1 phase by degrading the degrader of G1-S cyclins, and its absence results in G2 arrest \cite{13nc,11jtbi}. Moreover, our results also provide some new predictions. For example, we predict that increasing the binding strength for CycB and Cdk1 (parameter 24) will promote the cell cycle. This is probably because that CycB acts as a factor promoting the cell cycle. Also, we reveal that the synthesis of AP1 (parameter 1 and 14) hinders the crossing through the M checkpoint. These predictions can be tested further in experiments and provide potential drug targets related with the cell cycle.

It is believed that the accelerated pace of the cell cycle is related with cancer progression. A possible reason is that mutations or environmental factors lead to diminished landscape barriers and a compressed potential landscape gradient along the cell cycle trajectory, resulting in uncontrolled cycling. This imbalance can be addressed by modulating genetic or regulatory pathways through targeted interventions or environmental adjustments. Guided by the findings of our global sensitivity analysis, such perturbations may offer promising strategies for restoring stability of the system and mitigating cancer progression.

\section*{Discussion}

In this study, we introduce an approximation approach, named as DDGA, for quantifying the energy landscape of high-dimensional oscillatory networks. We applied this approach to various complex biological networks with diverse dimensions, displaying its high efficiency and effectiveness for high dimensionality. To conquer the inefficacy of the original approximation approaches, our approach involves two steps. First, we provide a pre-solution, which constructs a distribution on the limit cycle, to capture the one-dimensional stochastic dynamics and provide a comprehensive overview of the oscillation structure. Second, we incorporate diffusion effects based on the WSGA framework. We establish that the diffusion process on the orthogonal normal plane of the limit cycle can be approximated as a stationary process, simplifying a majority of ODEs into matrix equations. As a result, the computation for the covariance matrix is substantially expedited. Through diffusion decomposition, our DDGA dissects the stochastic evolution into the tangential and the orthogonal directions, effectively reconciling the trade-off between including non-diagonal covariance terms and avoiding ``covariance explosion''. 

To showcase the advantages of the DDGA, we evaluate its performance on two distinct oscillatory systems with dimensions 6 and 44. Through two probability measure indices and time-cost analysis, we benchmark the DDGA against two widely-used methods, the WSGA and the EGA. The results suggest that the DDGA facilitates the improved quantification of the landscape for the periodic oscillatory systems, and also promotes the explanations of intricate biological mechanisms. For example, it identifies the explosion phenomenon in a 6-dimensional synthetic oscillatory network and detects new basins and checkpoints that previous approaches fail to discern in a 44-dimensional mammalian cell cycle network. The landscape constructed by the DDGA can better explain the biological phenomena observed in experiments, such as the stability and the sensitivity of the oscillation, and agrees with single cell experimental data. Our approach offers a valuable tool for studying stochastic and periodic dynamics, as well as promoting mechanistic understanding of biological oscillatory systems, by quantifying associated energy landscape.

It should be noted that FPE poses significant challenges when approximating solutions for high-dimensional and nonlinear systems. Recent studies have also addressed these difficulties using trajectory density distribution methods and deep learning approaches \cite{Zhao2024EPR, VU2024CNSNS, Zhang2024Arxiv, Lin2023JCP}. For instance, Zhao et al. \cite{Zhao2024EPR} constructed loss functions based on the FPE operator and the invariant distribution, which is approximated using trajectory density distribution, to learn the landscape of the system. These approaches are highly versatile and effective for general complex systems. However, their numerical performance is sensitive to varying hyperparameters. Our DDGA approach, as an efficient and accurate approximation, may provide a robust preconditioner to facilitate the convergence of these methods. Thus, the combination of these two types of approaches holds significant potential to improve the reliability of solving high-dimensional oscillatory systems.

On the other hand, investigating underlying mechanisms from empirical data is a key challenge in biological systems. Researchers need reliable methods to quantify the landscapes from data for further analysis. While the DDGA method can effectively construct accurate energy landscapes from dynamic models for oscillatory systems, it cannot be directly applied to empirical data without combining it with methods that recover the system's underlying dynamics from observed data. Recent studies have introduced promising techniques for this purpose \cite{Cui2023MBR, Cui2023PM, Lin2023JCP, Zhou2024isci}. Integrating DDGA with these methods may provide a powerful tool for constructing landscape from experimental data, which warrants further explorations.

\section*{Supplemental Information}

Further Supplemental Information can be seen in https://doi.org/10.1016/j.xcrp.2025.102405

\normalem
\bibliographystyle{CRPS}

%%% the following code can also be used to create the references without .bib

\end{document}